\documentstyle[twocolumn,seceq]{jpsj}
\renewcommand{\theequation}{\arabic{section}.\arabic{equation}}
\catcode`\@=11
\def\simle{\mathrel{\mathpalette\@versim<}}   % < over \sim
\def\simge{\mathrel{\mathpalette\@versim>}}   % > over \sim
\def\@versim#1#2{\lower2.5pt\vbox{\baselineskip0pt \lineskip-.5pt
   \ialign{$\m@th#1\hfil##\hfil$\crcr#2\crcr\sim\crcr}}}
\catcode`\@=12
\def\runtitle{
Magnetization Process of Nanoscale Iron Cluster
}
\def\runauthor{Hiroki {\sc Nakano} and Seiji {\sc Miyashita}}

\title
{
Magnetization Process of Nanoscale Iron Cluster
}

\author
{ 
Hiroki {\sc Nakano} and Seiji {\sc Miyashita}
}

\inst
{
Department of Applied Physics, School of Engineering, 
University of Tokyo, \\ 
Hongo 7-3-1, Bunkyo-ku, Tokyo 113-8656\\
}

\recdate
{
\today
}

\abst
{
Low-temperature magnetization process of 
the nanoscale iron cluster in linearly sweeped fields 
is investigated by a numerical analysis 
of time-dependent Schr$\ddot{\rm o}$dinger equation and 
the quantum master equation.  
We introduce an effective basis method 
extracting important states, 
by which we can obtain the magnetization process effectively. 
We investigate the structure of the field derivative 
of the magnetization.  
We find out that the antisymmetric interaction determined 
from the lattice structure reproduces well 
the experimental results of the iron magnets and that  
this interaction plays an important role in the iron cluster. 
Deviations from the adiabatic process are also studied. 
In the fast sweeping case, our calculations indicate that 
the nonadiabatic transition dominantly occurs  
at the level crossing for the lowest field.  
In slow sweeping case, due to the influence 
of the thermal environment to the spin system, 
the field derivative of the magnetization shows 
an asymmetric behavior, the magnetic F$\ddot{\rm o}$hn effect, 
which explains the substructure of the experimental results 
in the pulsed field. 
}

\kword
{
Nanoscale magnet, magnetization process, 
time-dependent Schr$\ddot{\rm o}$dinger equation,
quantum master equation, 
Lanczos diagonalization, Runge-Kutta method, 
nonadiabatic transition
}

\begin{document}
\sloppy
\hyphenation{Hamil-ton-ian}
\maketitle

\section{Introduction}
Nanoscale magnets have attracted much interest 
from the point of view of their quantum dynamics.\cite{Caneschi_rev} 
As such magnetic materials, Fe$_n$ ($n=6,10,12,18$) 
have been synthesized. 
In these compounds, each iron ion Fe$^{3+}$ plays a role of 
an $S=5/2$ spin and forms a ring-structure molecule. 
Each of these compounds reveals a step-like magnetization process 
$M(H)$ at low temperatures. 
These steps appear due to the fact 
that the energy levels are discrete, 
which is one aspect of the quantum nature. 
First, 
let our attention be focused on the magnetizations differentiated 
by the magnetic field, 
${\rm d}M/{\rm d}H$\cite{comment_expdM-dH}. 
At magnetic fields where the magnetization grows rapidly, 
${\rm d}M/{\rm d}H$ shows peaks with various heights. 
In two of the compounds 
[M Fe$_6$ (OCH$_3$)$_{12}$(dbm)$_6$]$^+$ 
(M=Na,Li)\cite{Caneschi_Fe6} and 
[Fe(OMe)$_2$(O$_2$CCH$_2$Cl)]$_{10}$\cite{Taft_Fe10}, 
, which are abbreviated by Fe$_6$ and Fe$_{10}$ respectively,
the peak at the lowest magnetic field 
has the largest height; the heights of the peaks 
get smaller gradually at higher fields. 
How does the difference of the peak heights appear? 

The main structure of the magnetization process is 
determined by the Heisenberg-like exchange interactions 
between the iron ions, where the $z$-component of the 
magnetization $M_z$ is a good quantum number.  
Eigenstates of various values of $M_z$ show 
many crossings in the energy-level structure  
as function of the magnetic field $H$.  
The ground state is given by the lowest eigenstate 
at the given field, which changes piecewise.  
The typical case of antiferromangetc interactions is shown 
schematically in Fig.~\ref{fig1}. 
The singlet ground state, which is nonmagnetic, 
is realized at $H=0$.  
At $H=H_1$ and $H_2$,  
the ground state changes, where 
the magnetization of the ground state increases.  
If a system includes only the Heisenberg-like exchange interactions, 
${\rm d}M/{\rm d}H$ reveals the $\delta$ function at the crossing. 
In experiments, however, they have finite peaks.  
There are two possible origins for the finite heights. 
One is an effect due to finite temperature. 
The other is an effect due to some quantum-mechanical operators.  
Let us consider the case of the thermal effect. 
At low temperatures, 
only the two eigenstates concerning the crossing are relevant.  
Main contribution in the partition function 
consists of the Boltzmann weights of these states,  and 
the thermal behavior is characterized by the energy difference 
between them. This effect causes almost equivalent peaks 
at all the crossings in ${\rm d}M/{\rm d}H$, 
which is not the case in the experiments.  
Therefore, it seems that the origin of the difference 
between the peak heights is due to some operators 
which do not commute with $M_z$. 
What kind of operator does induce such a transition? 
In this paper, we investigate such interactions.  
We consider the lattice structure of the iron clusters and 
find the existence of the antisymmetric interaction, 
namely the Dzyaloshinsky-Moriya interaction.  
This interaction which does not commute with $M_z$ 
is found to reproduce well the feature observed experimentally 
in the magnetization process of the iron clusters.  

It is known that when the parameter is swept 
crossing the transition point, the system shows the nonadiabatic 
transition \cite{Landau,Zener,Stukelberg}.  
The transition rate depends on the sweeping speed 
of the field. 
In particular, the present case belongs to the so-called 
Landau-Zener-St$\ddot{\rm u}$kelberg (LZS) mechanism. 
Tunneling phenomena of the metastable magnetization 
of the Ising-type model were studied 
in the viewpoint of the nonadiabatic transition\cite{Miyashita95}. 
Here we treat a model for the nanoscale magnets and 
clarify how effects of the nonadiabatic transition appear.  

The peaks in ${\rm d}M/{\rm d}H$ of Fe$_{6}$ and Fe$_{10}$ 
are known to be asymmetric in higher- and lower-field regions 
with respect to the center of each peak.  
In the receding side of the pulsed field, the slope of the peak 
in ${\rm d}M/{\rm d}H$ is gentle 
in comparison with that of the approaching side.  
In the extreme case, 
a satellite peak is accompanied in the receding side.  
Similar behavior of hysteresis in the magnetization process 
is known in [Fe(salen)Cl]$_2$ (so-called Fe$_2$)\cite{Shapira_Fe2} 
and K$_6$[V$_{15}$As$_6$O$_{42}$(H$_2$O)] 
(so-called V$_{15})$\cite{V15_Chiorescu}. 
The behavior of the vanadium compound was investigated 
as influence of the thermal environment 
which is called magnetic F$\ddot{\rm o}$hn 
effect\cite{Mag_Foehn_effect}.  
In this work, we investigate 
the magnetization process of Fe$_6$ 
with the effect of environment 
using the quantum master equation. 
We find out that the asymmetry of the peaks 
of the iron clusters can be understood 
as an effect from the environment.  

This paper is organized as follows. 
The next section is devoted to the model Hamiltonian and 
the calculation method. 
The antisymmetric interaction in the model is introduced, 
which explains the main structure of the peaks 
in ${\rm d}M/{\rm d}H$.  
As to the method, 
we introduce an effective basis method in 
a dynamical approach in the Schr$\ddot{\rm o}$dinger equation
and also in the quantum master equation. 
In \S 3 the magnetization process at zero temperature 
is discussed. 
In \S 4 the magnetization process at finite temperatures 
is investigated.  
Concluding remarks are given in \S 5.  

\section{Hamiltonian and Method}

\subsection{Spin Hamiltonian}

We consider a system that has a large interaction 
of Heisenberg type between two neighboring spins.  
In addition, there presumably exists 
other anisotropic interactions determined 
from the lattice structure of the compound 
even though they are smaller than the Heisenberg-type one.  
Let us consider the lattice structure of the iron clusters.  
In both Fe$_6$ and Fe$_{10}$, two neighboring iron ions 
have a mirror symmetry with respect to the plane perpendicular 
to the line connecting these iron ions. 
In this case, it is known that 
the Dzyaloshinsky-Moriya (DM) interactions
\begin{equation}
{\cal H}_{\rm DM}=\sum_{j=1}^{N_{\rm s}} 
\mbox{\boldmath $D$}_j \cdot 
[\mbox{\boldmath $S$}_j \times \mbox{\boldmath $S$}_{j+1}],
\label{DM_interactions}
\end{equation}
may be present, and that $\mbox{\boldmath $D$}_j$ is 
parallel to the mirror plane.  
Here, $N_{\rm s}$ denotes the number of sites and  
$\mbox{\boldmath $S$}_j$ denotes an $S=5/2$ spin operator 
at site $j$.  
The system is periodic, i.e., 
$\mbox{\boldmath $S$}_{N_{\rm s}+1}=\mbox{\boldmath $S$}_1$.  
It is known that, in every two neighboring pairs, 
branches of atomic bonds are alternating.  
Therefore, $\mbox{\boldmath $D$}_j$ and 
$\mbox{\boldmath $D$}_{j+1}$ have opposite directions.  
In this work, we treat the regular hexagon ($N_{\rm s}=6$)
with alternating $\mbox{\boldmath $D$}_j$ 
\begin{equation}
\mbox{\boldmath $D$}_j=(-1)^{j} (D,0,0), 
\label{D_in_DM}
\end{equation}
as shown in Fig.~\ref{fig2}.  
Thus, as a candidate model for the system of Fe$_6$ and Fe$_{10}$, 
we examine the Hamiltonian 
\begin{equation}
{\cal H}_{\rm sys}={\cal H}_0+{\cal H}_{\rm DM},
\label{total_Hamiltonian}
\end{equation}
\begin{equation}
{\cal H}_0=\sum_{j=1}^{N_{\rm s}} [J 
\mbox{\boldmath $S$}_j \cdot \mbox{\boldmath $S$}_{j+1}
- h (t) S_j^z ], 
\label{isotropic_term}
\end{equation}
where $J$ is an amplitude of antiferromagnetic 
exchange interaction, and  
$h(t)$ is the external magnetic field.  
Here we consider the case of linear sweeping, $h(t)=ct$.  
In this work we examine the case where 
$D$ is much smaller than $|J|$.  
A main structure of energy is determined 
by ${\cal H}_{\rm 0}$ as a function of the field, 
as shown schematically in Fig.~\ref{fig1}.  
Because $J \gg D$, 
the effect of ${\cal H}_{\rm DM}$ appears 
only in the vicinity of a level crossing, and 
causes the avoided level crossing structure.  

\subsection{Calculation method}

Let us obtain the magnetization process.  
In this paper, we introduce a dynamical approach to obtain 
the magnetization process.  
It is possible in principle to calculate 
the adiabatic magnetization process $M(h)$ 
of ${\cal H}_{\rm sys}$ by diagonalization.  
However, in order to obtain smooth curves, 
very fine steps in the field are necessary, 
which requires a large number of diagonalizations.  
Moreover this dynamical approach provides information 
on the dynamical processes which cannot be obtained 
by diagonalization.  

To obtain the magnetization process in the ground state ($T=0$), 
we solve the time-dependent Schr$\ddot{\rm o}$dinger equation 
\begin{equation}
{\rm i}\hbar \frac{\rm d}{{\rm d}t} |\psi \rangle = 
{\cal H}_{\rm sys} |\psi \rangle ,
\label{org_schrodinger_eq}
\end{equation}
numerically. 
This equation is made dimensionless 
by the scaling $\tilde{t}=Jt/\hbar$.  
We take $J$ as an energy unit; 
$\tilde{h}=h/J$ and $\tilde{D}=D/J$ are independent parameters 
in the Schr$\ddot{\rm o}$dinger equation. 
The external magnetic field $\tilde{h}$ is increased linearly 
from zero, which corresponds to the pulsed field in experiments.  
In particular, although the pulse velocity has 
a sinusoidal shape in experiments, 
here we approximate it by a constant velocity.  
As far as the velocity is small, 
we can obtain the adiabatic process in both cases.  
The antiferromagnetic ground state for $h=0$ 
is taken as an initial state.  

To treat the thermal environment, 
we have to consider the effect of environment, 
where the total Hamiltonian is given by 
\begin{equation}
{\cal H}_{\rm tot}={\cal H}_{\rm sys}+{\cal H}_{\rm int}
+{\cal H}_{\rm bath},
\end{equation}
where ${\cal H}_{\rm bath}$ 
denotes a Hamiltonian of the heat bath. 
${\cal H}_{\rm int}$ denotes a term 
of interaction, which plays a role of interface 
exchanging between the heat bath and the spin system. 
We take the heat bath as the free boson system, namely, 
${\cal H}_{\rm bath}=\sum_{\alpha} \hbar \omega_{\alpha}
(b^{\dagger}_{\alpha} b_{\alpha} +1/2 )$.  
${\cal H}_{\rm int}$ is taken as 
${\cal H}_{\rm int}=\lambda \sqrt{\hbar} 
\sum_{\alpha} (b^{\dagger}_{\alpha}+ b_{\alpha}) X
$, where $X$ is an operator in the spin system. 
Here, $\langle \phi_{M} | X |\phi_{M^{\prime}}\rangle$ is taken  
as a unit when $|M-M^{\prime}|=1$ and vanishes otherwise.  
Under the condition of small coupling $\lambda$, the quantum master equation 
for the density matrix 
\begin{equation}
\frac{d \rho (t)}{dt}=\frac{1}{{\rm i}\hbar}
[{\cal H}_{\rm sys},\rho (t)] - \Gamma \rho (t), 
\label{quantum_master_eq}
\end{equation}
has been derived\cite{QME}.  
Here, 
\begin{equation}
\Gamma \rho (t)=\{[X,R \rho (t)]+[X,R \rho (t)]^{\dagger}\}
\lambda^2 \pi/\hbar, 
\end{equation}
\begin{equation}
\langle k | R | n \rangle = \frac{1}{\hbar}X_{k,n}
\Phi\left(\frac{E_k-E_n}{\hbar}\right),
\end{equation}
\begin{equation}
\Phi(\omega)=\frac{I(\omega)-I(-\omega)}{{\rm e}^{\beta\omega}-1},
\end{equation}
where $\beta$ is the inverse of temperature, 
namely $k_{\rm B} T=1/\beta$. 
Here, $|k\rangle$ and $|n\rangle$ represent eigenstates 
for ${\cal H}_{\rm sys}$ with energy eigenvalues 
$E_{k}$ and $E_{n}$, respectively. 
If we assume that a main contribution of the heat bath 
comes from the lattice, the bosons in the environment 
are considered to be phonons.  
Then, the spectral density $I(\omega)$ is given 
by $I_0 \omega^2 \theta(\omega)$ 
where  $\theta(\omega)$ is the step function.  
We solve the quantum master equation 
(\ref{quantum_master_eq}) to obtain the magnetization process 
at finite temperatures.  

\subsection{Effective basis method}

Since the number of states 
$N_{\rm D}=(2S+1)^{N_{\rm s}}$=46656 is large, 
solving the Schr$\ddot{\rm o}$dinger equation 
(\ref{org_schrodinger_eq}) using all the states 
in the system is difficult, in particular for long times.  
The calculation of the quantum master equation 
(\ref{quantum_master_eq}) is even more difficult because 
the number of elements of the density matrix is $N_{\rm D}^2$.  
Therefore, some approximation method 
which extracts important states of the system is required. 

Now we introduce an approximate method for this purpose.  
First note here that many eigenstates are irrelevant 
to study the magnetization growth with the increasing field 
in the ground state through the quantum transitions 
of the present interests. 
In spite of this fact, 
the irrelevant states and important states are 
treated equivalently in full-size calculations, 
and this is inefficient.  
Now let us consider a method to extract the important states 
which contribute to the quantum transitions.  

Here, the following procedure is performed, which we call 
an effective basis method.   
At first, one obtains the ground state $|\psi_{\rm G}\rangle$ 
of ${\cal H}_{\rm sys}$ for a field $\tilde{h}(\tilde{t})$ 
at a time $\tilde{t}$ 
(initially, $\tilde{t}=0$) by the Lanczos diagonalization 
in the basis taking the $z$ direction as the quantization axis.  
Note that for the present system with dimension $N_{\rm D}=46656$, 
this is still possible.   
From $|\psi_{\rm G}\rangle$, one extracts 
$ 
|\phi_{M_z}\rangle=|\tilde{\phi}_{M_z}\rangle /
\sqrt{\langle\tilde{\phi}_{M_z}|\tilde{\phi}_{M_z}\rangle},
$ where 
\begin{equation}
|\tilde{\phi}_{M_z}\rangle=\sum_{k}
|\xi_{k}\rangle
\langle\xi_{k}|\psi_{\rm G}\rangle ,
\label{extraction}
\end{equation}
where $|\xi_{k}\rangle$ denotes the $k$-th basis state with 
the magnetization $M^z$, i.e., 
$S_z^{\rm tot}|\xi_k\rangle=M^z |\xi_k\rangle$.  
Here, $S_z^{\rm tot}$ is defined by 
$\sum_{j=1}^{N_{\rm s}} S^z_j$. 
Note that $M_z$ is an integer ranging 
between zero and the saturation value, which is equal to 15 
(=$\frac{5}{2}\times 6$) for the present case. 
The new state $|\phi_{M_z}\rangle$ is a representative 
of the states of $M_z$.  
We have checked that the set of states $\{|\phi_{M_z}\rangle\}$ 
represents well the behavior of ${\cal H}_{\rm sys}$ at low energy.  
Using this set of states, one constructs 
a small matrix 
\begin{eqnarray}
& &\tilde{\cal H}_{\rm sys}= \nonumber \\
& &
\left[\begin{array}{cccc}
  \langle\phi_{0}|{\cal H}_{\rm sys}|\phi_{0}\rangle 
& \langle\phi_{0}|{\cal H}_{\rm sys}|\phi_{1}\rangle 
& \cdots  
& \langle\phi_{0}|{\cal H}_{\rm sys}|\phi_{15}\rangle \\
  \langle\phi_{1}|{\cal H}_{\rm sys}|\phi_{0}\rangle 
& \langle\phi_{1}|{\cal H}_{\rm sys}|\phi_{1}\rangle 
& \cdots 
& \langle\phi_{1}|{\cal H}_{\rm sys}|\phi_{15}\rangle \\
  \vdots
& \vdots
& \ddots
& \vdots \\
  \langle\phi_{15}|{\cal H}_{\rm sys}|\phi_{0}\rangle 
& \langle\phi_{15}|{\cal H}_{\rm sys}|\phi_{1}\rangle 
& \cdots 
& \langle\phi_{15}|{\cal H}_{\rm sys}|\phi_{15}\rangle \\
\end{array}
\right]. \nonumber \\
& & 
\end{eqnarray}
Here, we take only the ground state to generate 
the effective basis.  
In this paper, we find that even in the present basis, 
we obtain good results.  
However, we can improve the approximation by adding more 
basis states extracted from a few low-energy states. 

In this work, 
we perform Runge-Kutta calculations 
for the small matrix $\tilde{\cal H}_{\rm sys}$ 
instead of the full-size matrix ${\cal H}_{\rm sys}$.  
In the Runge-Kutta calculations, the normalization 
of the wave function deviates easily from unity.  
We have checked it carefully and used a small enough time step.  
The initial condition is $|\phi_{0}\rangle$.   
To keep the quality of the approximation the same, 
we update a set \{$|\phi_{M_z}\rangle$\} 
by the above mentioned method, i.e. (\ref{extraction}),  
after time evolution by $\cal{H}$ for a while.  
It is found that \{$|\phi_{M_z}\rangle$\} gives a good approximation 
if it is updated as often as the level crosses.  
We have verified that no significant differences appear 
when the frequency of the update is varied. 
When the sweeping speed is small and the system 
behaves adiabatically, 
the states with the lowest and the second lowest energies 
are important.  
The method described here gives 
the magnetization process $M(h)$ 
as the adiabatic process very effectively.  
Actually, this dynamical method is a natural way corresponding 
to experiments of pulsed magnetic field.  

As we mentioned, the present method provides information 
of nonadiabatic process for fast sweepings, 
that cannot be obtained by a static Lanczos diagonalization only. 
To demonstrate the validity of the above approximation 
for the nonadiabatic transition, 
we have performed the calculation 
of the Schr$\ddot{\rm o}$dinger equation for $S=1$ 
for $\tilde{D}=0.01$.  
In this case, the first avoided level crossing has an energy gap 
$\Delta E /J \sim 0.0196$.  
According to the LZS mechanism, the transition rate 
at an avoided crossing is given by
\begin{equation}
P=1-\exp\left(-\frac{\pi (\Delta E/J)^2}{2 |M_z-M_z^{\prime}| 
\tilde{h}_{\rm max}/\tilde{t}_{\rm max}}\right) , 
\label{LZS_rate}
\end{equation}
when the magnetic field is linearly sweeped from zero 
to $\tilde{h}_{\rm max}$ during a time of $\tilde{t}_{\rm max}$. 
For the above energy gap, $\tilde{h}_{\rm max}=6$, 
$\tilde{t}_{\rm max}=3000$, $M_z=1$ and $M_z^{\prime}=0$,  
the transition rate $P$ given by (\ref{LZS_rate}) is $0.26$. 
After the corresponding level crossing, 
the present approximation gives for this rate 0.26. 
This agreement indicates the validity of the above approximation.  
For the quantum master equation, we use the same effective basis.  
We have also checked the agreement of a result 
from the quantum master equation for a full-size Hamiltonian 
with a small dimension and 
that from the quantum master equation for the Hamiltonian 
with its effective basis, 
which also indicates the validity of the approximation 
in the quantum master equation. 

\section{Magnetization Process at $T=0$}

Figure~\ref{fig3}(a) shows the result of the magnetization process 
for $\tilde{t}_{\rm max}=30000$, 
$\tilde{h}_{\rm max}=15$ and $\tilde{D}=0.01$ together 
with the complete stair case of the ground state 
of the model of $D=0$.  
Every step indicates the position of a level crossing 
at which the magnetization grows from $M$ to $M+1$.   
Because a final value of the magnetization 
reaches full saturation, 
the process is almost adiabatic at every crossing.  
Figure~\ref{fig3}(b) shows ${\rm d}M/{\rm d}\tilde{h}$ 
for $\tilde{D}=0.01$.  
Peaks with various heights induced by the DM interaction 
are formed at the level-crossing positions denoted by crosses. 
It is noticable that the heights of the peaks 
are different from each other. 
Especially, the first peak is the highest.  
The heights gradually become small as the 
field is increased 
although the heights in a few last peaks 
get higher again.  
These features of the different peak heights agree well 
with the experimental results in ${\rm d}M/{\rm d}H$ 
for Fe$_6$\cite{Caneschi_Fe6} and Fe$_{10}$\cite{Taft_Fe10}. 

Single-ion-type anisotropy and 
dipole-dipole interactions may causes broadening 
of ${\rm d}M/{\rm d}H$.  
However, the contribution of the single-ion anisotropy is considered 
to be small because the orbital angular momentum vanishes 
in the each iron ion Fe$^{3+}$ of the above materials.  
The off-diagonal element  
$\langle \phi_{2M+1} | {\cal H}_{\rm dip} |\phi_{2M}\rangle$ 
of the dipole-dipole interactions, 
\begin{equation}
{\cal H}_{\rm dip}\equiv\sum_{i,j}
(g\mu_{\rm B})^2[
\frac{\mbox{\boldmath $S$}_i\cdot
\mbox{\boldmath $S$}_j}{r_{ij}^3 }
- \frac{3}{r_{ij}^5} 
(\mbox{\boldmath $S$}_i\cdot\mbox{\boldmath $r$}_{ij})
(\mbox{\boldmath $S$}_j\cdot\mbox{\boldmath $r$}_{ij})
],
\end{equation}
vanishes 
because $|\phi_{2M}\rangle$ and $|\phi_{2M+1}\rangle$ have 
wave numbers $\pi$ and 0, respectively. 
This means that 
contribution of the dipole-dipole interactions is also small.  
Thus, we conclude that the transition 
due to the alternating DM interactions 
is the major origin of the peaks of ${\rm d}M/{\rm d}H$ 
of Fe$_6$ and Fe$_{10}$ with different heights. 

On the contrary the ${\rm d}M/{\rm d}H$ curve of 
Fe$_{12}$, which is an abbreviation 
of [Fe(OCH$_3$)$_2$(dbm)]$_{12}$\cite{Caneschi_Fe12}, 
reveals that the height of the first peak in ${\rm d}M/{\rm d}H$ is 
not the largest and that the second one is larger 
than the first one\cite{Ajiro_Narumi_private}. 
It was reported that the lattice structure of this material has 
a large deviation from the regular polygon, 
while in Fe$_6$ and Fe$_{10}$, 
the lattices are almost regular polygons.  
In order to take the deviation into account, we study 
a lattice with locally non-uniform magnetic interaction \{$J_i$\}.  
In our calculations, however, the first peak is still the highest 
in the system with various sets of \{$J_i$\}.  

In the dimer material Fe$_2$, the second peak 
is higher than the first one\cite{Shapira_Fe2}.  
The interactions in these compounds 
should be clarified in the future.  

So far we studied the adiabatic behavior for slow sweeps, 
which probes the magnetization processes 
in the ground state.  
Next, let us consider the dynamical effect of the sweeping 
of the magnetic field in the case of a larger speed.  
The result for $\tilde{t}_{\rm max}=7500$ 
is shown in Fig.~\ref{fig4}.  
The magnetization at a field 
where the saturation value is reached 
in Fig.~\ref{fig3}(a) becomes smaller.  
This means that there are some level crossings at which 
nonadiabatic transitions occur. 
At $\tilde{h}=0.7$, the magnetization jumps up to 
$M=1$ in the adiabatic process.  
However, in Fig.~\ref{fig3}(a), 
the step height is clearly lower, 
suggesting a nonadiabatic transition at the first level crossing.  
In the fields after the first crossing before the second one, 
the state is given approximately by a linear combination 
of $|\phi_0\rangle$ and $|\phi_1\rangle$, 
i.e., $|\psi\rangle=c_0 |\phi_0\rangle+c_1 |\phi_1\rangle$, 
where $|\phi_0\rangle$ and $|\phi_1\rangle$ behave 
almost as eigenstates for the field away from the crossings. 
The probabilities of 
$|\phi_0\rangle$ and $|\phi_1\rangle$ are found to be 
$|c_0|^2 = 0.22$ and $|c_1|^2 = 0.78$, respectively. 
The plateau height $m$ satisfies the relation 
$m=|c_1|^2=1-|c_0|^2=0.78$.  
For fields between the second and 
the higher-field crossings, 
the growth of the magnetization is briefly linear. 
The population of $|c_1|^2 = 0.78$ stays in the ground state, 
which gives the linear increase of the magnetization.  
The saturated magnetization 
for $\tilde{t}_{\rm max}=7500$ agrees with 15$\times m$. 
These facts indicate 
that the system shows almost adiabatic process 
at crossings other than the first one 
within this sweeping speed and 
that the energy gap at the first crossing is found to be the smallest. 
Thus, the transition at the first crossing 
plays a role of a bottleneck process 
for the growth of the magnetization. 
Note here that 
the population at $|\phi_0\rangle$ does not cause changes 
in the magnetization process within the present approximation 
although it is also exposed to nonadiabatic transitions 
in higher fields in the full-size Hamiltonian.  
To obtain such nonadiabatic transitions, 
an improvement of the method of calculation is required.  

If the nonadiabatic transitions are experimentally observed 
as a quantum effect, 
we could capture the phenomena of 
the states which branch at the level crossings.  
Duration times 
$\tilde{t}_{\rm max}=7500$ and $\tilde{t}_{\rm max}=30000$ 
correspond to 3$\sim$6 [ns] and 11$\sim$22 [ns] 
in real time, respectively, 
by using $J/k_{\rm B} = 10\sim 20$~[K] estimated  
from the susceptibility fit of Fe$_6$ and Fe$_{10}$.  
In the present experiments, on the other hand, 
the duration time of the pulsed field is in order of 10 [ms].  
In order to find the nonadiabatic phenomena 
in Fe$_6$ and Fe$_{10}$, either a faster sweeping speed or 
a value of $\tilde{D}$ smaller than 0.01 is necessary.  
When $\tilde{D}$ is small, 
it is expected that 
$D$ is proportional to the energy gap at the level crossing. 
The amplitude of the gap determines whether the transition 
is adiabatic or nonadiabatic.  
Then, for the appearance of the nonadiabatic features 
$D/k_{\rm B}$ must be in order of 0.1 [mK], 
if the duration time is 10 [ms] and $J/k_{\rm B}=10$ [K].  
An experimental estimate of the amplitude 
of the antisymmetric interactions in Fe$_6$ and Fe$_{10}$ 
would be very helpful.  

\section{Magnetization Process at Finite Temperatures}

In the previous subsection, 
we studied the dynamical magnetization process  
in the pure quantum-mechanical way.  
At finite temperatures, however, 
influence from the surrounding environment 
may have significant effects.  
Here, we study the effect of the thermal environment 
on the iron clusters.  

Figure~\ref{fig5} shows the result of the magnetization process 
for the case of $\tilde{D}=0.01$, $\lambda=0.005$, 
$\tilde{h}_{\rm max}=12$ and $\tilde{t}_{\rm max}=60000$ 
in the condition of temperature $k_{\rm B} T/J=0.05$.    
A satellite peak in ${\rm d}M/{\rm d}\tilde{h}$, 
is clearly observed around $\tilde{h}=0.9$ 
in the receding side of the field 
near the main peak at $\tilde{h}=0.7$.  
This satellite peak corresponds to a dent 
of the magnetization process at an edge of the magnetization step.  
This satellite peak is considered to be a consequence 
of the magnetic F$\ddot{\rm o}$hn effect\cite{Mag_Foehn_effect}.  
The key mechanism of the magnetic F$\ddot{\rm o}$hn effect is 
the difference between two kinds of speed 
characterizing properties of the total system. 
One is a speed determined by the LZS mechanism, 
below which the quantum transition is almost adiabatic.  
The other is a speed controlled by ${\cal H}_{\rm int}$, 
above which the supply of energy from the heat bath 
is not fast enough to keep the system in the equilibrium 
during the change of the field.  
If the sweeping speed of the field is between these two, 
the magnetic F$\ddot{\rm o}$hn effect occurs.  
This effect has been proposed as a general phenomenon 
in the processes which are almost adiabatic 
but with a little inflow of heat\cite{Mag_Foehn_effect}. 
Because the process is almost adiabatic, 
the system is almost in the ground state, 
that is in the lower-energy state.  
However, due to the interaction with the environments, 
some population is excited to the higher-energy level.  
But it is not enough to realize 
the thermal equilibrium distribution.  
This excitation occurs as far as the equilibrium population 
at the higher-energy level $\rho^{(2)}_{\rm eq} (\propto 
{\rm e}^{-\beta \Delta(\tilde{h})})$ is larger than 
the dynamical one of the system  $\rho^{(2)} (t)$, 
when $\Delta(\tilde{h})$ is the energy difference 
between the states.  
When the field gets away from the crossing point, 
the energy difference increases; 
$\rho^{(2)}_{\rm eq}$ decreases rapidly and 
becomes smaller than $\rho^{(2)} (t)$.  
There, the population at higher-energy level begins to relax 
to the lower-energy level. In this way, $\rho^{(2)} (t)$ has 
a maximum at a field $\tilde{h}$ in the receding side 
from the crossing point.  The behavior of $\rho^{(2)} (t)$ 
causes a dent of the magnetization process.  

In the receding side of the field 
near the main peak at $\tilde{h}=1.4$, 
a smaller swell appears around $\tilde{h}=1.6$.  
In this case, the satellite due to the thermal redistribution 
is being merged into the main peak.    
Situations of merging depend 
on the distance between the satellite and the center of the main peak, 
and 
on the width of the main peak.  
The former is determined by the details of the heat bath and 
those of the interface interaction.  
The latter is given quantum-mechanically 
by the energy gap at the avoided level crossing.  
In the present case, the merging behavior is induced because 
the width of this main peak becomes larger 
in higher fields. 
This effect causes a gentle slope in the receding side compared 
with that in the approaching side.  
Thus, the asymmetry of the experimental peak in the field derivative 
of the magnetization is understood as a consequence of 
the magnetic F$\ddot{\rm o}$hn effect induced by the thermal environment. 

Let us consider the temperature dependence of the satellite peak. 
Results are shown in Fig.~\ref{fig6}.  
As the temperature is decreased, the satellite peak gets smaller 
and almost vanishes at $k_{\rm B} T/J=0.01$. 
It is characteristic that the satellite peak disappears 
without merging with the main peak.  
This temperature corresponds to $0.09\sim 0.18$~K 
if we take a experimental estimation of $J/k_{\rm B}$ 
to be $ 10\sim 20$~K.  
This means that in the experiments for Fe$_6$ 
at such low temperatures, 
the satellite peaks in ${\rm d}M/{\rm d}H$ would disappear 
in the way we observed.  
A detailed comparison with such an experiment is required.  
 
\section{Concluding Remarks}

We have studied properties and the origin of 
the quantum transitions observed 
in the magnetization process of the nanoscale iron clusters 
at $T=0$ and at finite temperatures.  
We have proposed an approximate procedure 
to make the numerical analysis possible 
and obtain the magnetization process for the system 
where the dimension of the basis states is large 
and conventional method is hardly applicable.  
We have found that the antisymmetric interaction reproduces 
the characteristics of the heights of the peaks 
in ${\rm d}M/{\rm d}H$ at $T=0$.  
The effect of nonadiabatic magnetization process 
was also discussed. 
The first level crossing plays 
a role of the major branching point. 
At finite temperatures, the asymmetric behavior of the peaks 
in ${\rm d}M/{\rm d}H$ is induced by the thermal environment
influencing the spin system as the magnetic F$\ddot{\rm o}$hn effect.  
In experiments, observation of the magnetization process is 
a powerful and direct method to estimate the coupling amplitude $J$ 
of the Heisenberg-like exchange interactions.  
However, the magnetization process yeilds 
not only information on $J$ 
but also other information characterizing the properties 
of the materials.  
In this paper, we have clarified such aspects 
of the iron clusters.  

\section*{Acknowledgements}
We would like to thank K.~Saito 
for valuable discussions.  
We also thank Y.~Narumi, Y.~Ajiro, Y.~Inagaki
and K.~Kindo 
for providing us 
with the experimental data before publication and 
for stimulating discussions.  
We wish to thank H.~De~Raedt for his kind critical reading 
of the manuscript.  
This work is supported partly by a Grant-in-Aid 
for Scientific Research from the Ministry of Education, Science, 
Sports and Culture, Japan. 
One of the authors (H.N.) is supported 
by Research Fellowships of the Japan Society 
for the Promotion of Science for Young Scientists.
A part of the computations was performed using the
facilities of the Supercomputer Center, 
Institute for Solid State Physics, University of Tokyo.


\begin{thebibliography}{99}
\bibitem{Caneschi_rev} A. Caneschi, D. Gatteschi, C. Sangregorio, 
R. Sessoli, L. Sorace, A. Cornia, M. A. Novak, C. Paulsen and 
W. Wernsdorfer: J. Magn. Magn. Mater. {\bf 200} (1999) 182 
and references therein. 
\bibitem{comment_expdM-dH} Experimentally, the magnetization 
is obtained from the integration of observed ${\rm d}M/{\rm d}H$.  
\bibitem{Caneschi_Fe6} A. Caneschi, A. Cornia, C. Fabretti, 
S. Foner, D. Gatteschi, R. Grandi and L. Schenetti: 
Chem. Eur. J. {\bf 2} (1996) 1379.
\bibitem{Taft_Fe10} K. L. Taft, C. D. Delfs, G. C. Papaefthymiou, 
S. Foner, D. Gatteschi and J. Lippard: 
J. Am. Chem. Soc. {\bf 116} (1994) 823.
\bibitem{Landau} L. Landau: Phys. Z. Sowjetunion {\bf 2} (1932) 46.
\bibitem{Zener} C. Zener: Proc. R. Soc. London Ser. A 
{\bf 137} (1932) 696.
\bibitem{Stukelberg} E. C. G. St$\ddot{\rm u}$kelberg: 
Helv. Phys. Acta {\bf 5} (1932) 369.
\bibitem{Miyashita95} S. Miyashita: J. Phys. Soc. Jpn. 
{\bf 64} (1995) 3207.
\bibitem{Shapira_Fe2} Y. Shapira, M. T. Liu, S. Foner, 
C. E. Dub$\acute{\rm e}$ and P. J. Bonitatebus Jr.: 
Phys. Rev. B {\bf 59} (1999) 1046.
\bibitem{V15_Chiorescu}
I. Chiorescu, W. Wernsdorfer, A. M$\ddot{\rm u}$ller, 
H. B$\ddot{\rm o}$gge and B. Barbara: 
Phys. Rev. Lett. {\bf 84} (2000) 3454. 
\bibitem{Mag_Foehn_effect}
K. Saito and S. Miyashita: cond-mat/0004027. 
\bibitem{QME}
K. Saito, S. Takesue and S. Miyashita: Phys. Rev. 
B {\bf 61} (2000) 2397. 
\bibitem{Caneschi_Fe12} A. Caneschi, A. Cornia, A. C. Fabretti 
and D. Gatteschi: Angew. Chem. Int. Ed. Eng. {\bf 38} (1999) 1295.
\bibitem{Ajiro_Narumi_private} 
Y.~Ajiro and Y.~Inagaki: private communications, 
Y. Narumi and K.~Kindo: private communications.
\end{thebibliography}
\end{document}